\documentclass[aps,prx,superscriptaddress,longbibliography,twocolumn,10pt]{revtex4-2}

\usepackage{graphicx}
\usepackage{amsmath}

\usepackage{amssymb}
\usepackage{hyperref}
\usepackage[utf8]{inputenc}
\usepackage{mathtools}
\usepackage[english]{babel}
\usepackage{bbm}
\hypersetup{colorlinks=true, linkcolor=blue, citecolor=blue, urlcolor=blue}
\usepackage{xcolor}
\usepackage{braket}
\usepackage{bbold}
\newcommand{\me}{\text{e}}
\newcommand{\iu}{\text{i}}
\usepackage[normalem]{ulem}
\usepackage{bbm}

\newcommand{\ii}{\mathrm{i}}

\newcommand{\dv}[1]{\frac{\text{d}}{\text{d}#1}}

\newcommand{\bea}{\begin{eqnarray}}
\newcommand{\eea}{\end{eqnarray}}

\binoppenalty=\maxdimen
\relpenalty=\maxdimen

\usepackage{pgf}

\usepackage{lineno}

\begin{document}

%\linenumbers

\title{Solving Dicke superradiance analytically: A compendium of methods}
\author{R.~Holzinger}
\email{raphael$_$holzinger@fas.harvard.edu}
\affiliation{Department of Physics, Harvard University, Cambridge, Massachusetts 02138, USA}
\affiliation{Institute for Theoretical Physics, University of Innsbruck, Technikerstraße 21a, 6020 Innsbruck, Austria}
\author{N.~S.~Bassler}
\affiliation{Max Planck Institute for the Science of Light, Staudtstraße 2, D-91058 Erlangen, Germany}
\affiliation{Department of Physics, Friedrich-Alexander-Universität Erlangen-Nürnberg, Staudtstraße 7, D-91058 Erlangen, Germany}
\author{J.~Lyne}
\affiliation{Max Planck Institute for the Science of Light, Staudtstraße 2, D-91058 Erlangen, Germany}
\affiliation{Department of Physics, Friedrich-Alexander-Universität Erlangen-Nürnberg, Staudtstraße 7, D-91058 Erlangen, Germany}
\author{F.~G.~Jimenez}
\affiliation{Pontificia Universidad Catolica del Peru, Av. Universitaria 1801, San Miguel 15088, Peru}
\author{J.~T.~Gohsrich}
\affiliation{Max Planck Institute for the Science of Light, Staudtstraße 2, D-91058 Erlangen, Germany}
\affiliation{Department of Physics, Friedrich-Alexander-Universität Erlangen-Nürnberg, Staudtstraße 7, D-91058 Erlangen, Germany}
\author{C.~Genes}
\email{claudiu.genes@physik.tu-darmstadt.de}
\affiliation{TU Darmstadt, Institute for Applied Physics, Hochschulstraße 4A, D-64289 Darmstadt, Germany}
\affiliation{Max Planck Institute for the Science of Light, Staudtstraße 2, D-91058 Erlangen, Germany}
\affiliation{Department of Physics, Friedrich-Alexander-Universität Erlangen-Nürnberg, Staudtstraße 7, D-91058 Erlangen, Germany}

\begin{abstract}
We present several analytical approaches to the Dicke superradiance problem, which involves determining the time evolution of the density operator for an initially inverted ensemble of $N$ identical two-level systems undergoing collective spontaneous emission. This serves as one of the simplest cases of open quantum system dynamics that allows for a fully analytical solution. We explore multiple methods to tackle this problem, yielding a solution valid for any time and any number of spins. These approaches range from solving coupled rate equations and identifying exceptional points in non-Hermitian evolution to employing combinatorial and probabilistic techniques, as well as utilizing a quantum jump unraveling of the master equation. The analytical solution is expressed as a residue sum obtained from a contour integral in the complex plane, suggesting the possibility of fully analytical solutions for a broader class of open quantum system dynamics problems.
\end{abstract}

\maketitle

%%%%%%%%%%%%%%%%%%%%%%%%%%%%%%%%%%%%%%%%%%%%%%%%%%%%%%%%%%%%%%%%%%%%

%%%%%%%%%%%%%%%%%%%%%%%%%%%%%%%%%%%%%%%%%%%%%%%%%%%%%%%%%%%%%%%%%%%%%

Dicke superradiance describes the rapid burst of radiation emitted by an initially fully inverted ensemble of $N$ two-level quantum systems (TLS) undergoing \textit{collective} decay~\cite{Dicke_originalpaper}. This is in stark contrast with the situation exhibited by a dilute, non-interacting, \textit{independently} emitting ensemble. To understand the emergence of such behavior, let us denote by $\Gamma$ the spontaneous emission rate of an isolated TLS from its excited state $\ket{e}$ to its ground state $\ket{g}$. The collapse operator associated with this loss process is denoted by $\sigma=\ket{g}\bra{e}$. For independent emitters, the rate at which the ensemble emits is identical to the one of an isolated system, so each constituent of the ensemble will lose excitation exponentially $\propto e^{-\Gamma t}$. Therefore, the time dependence of the total radiated power shows a trivial exponential dependence  $\propto N\Gamma e^{-\Gamma t}$. For the collectively decaying system, the collapse operator is $S=\sum_{j=1}^N \sigma_j$ and the decay rate turns out to be strongly time dependent. The fully inverted state decays at the same rate per particle as the isolated TLS, i.e. at rate $\Gamma$; however, the emergence of particle-particle correlations leads to an increase in the emission rate as the ensembles evolves toward half-occupancy states, to a maximum of $N\Gamma$ per particle. Meanwhile, as the total excitation number decreases from $N$ to zero, a peak of radiated power occurs when the system passes through the half-inverted state and the emission rate is close to the maximum rate of $N \Gamma$ per particle. This peak of total radiated power with amplitude $\propto N^2$ and occurring at short times $\propto \ln(N)/(N\Gamma)$ is the typical signature of superradiance~\cite{gross_haroche}.

The main aspect of Dicke superradiance is the permutation symmetry, owing to the symmetry of the collective collapse operator $S$. This feature was introduced by Dicke in 1954~\cite{Dicke_originalpaper} and analytically solved first by Lee in 1977, starting with the full excitation case~\cite{Tsung1977Part1} and generalizing to any arbitrary initial Dicke state~\cite{Tsung1977Part2}. Other analytical treatments followed, for example described in Refs.~\cite{Rupasov1984,Yudson1985}.\par
\indent The solution proposed in Refs.~\cite{Tsung1977Part1,Tsung1977Part2} has been mostly overlooked during the last decades. Presumably the convoluted analytical expression listed therein renders this solution computationally demanding, thus requiring resources similar to the direct simulation of the master equation. The focus instead shifted recently to derive more informative solutions in the mesoscopic (or thermodynamic) limit of large $N$, such as, for example, in \mbox{Refs.~\cite{approx1,approx2,approx3,approx4,molmer2021,Cirac2022LargeNLimit}}. A comprehensive early review of the problem and its implications can be found in Ref.~\cite{gross_haroche}.\par
\indent The original formulation of the Dicke superradiance model is hard to realize in practice, as closely positioned quantum emitters interact strongly (proportionally to $d^{-3}$, where $d$ is their separation), shifting the bare electronic levels and removing their indistinguishability. However, alternative realizations are available, for example in the context of cavity quantum electrodynamics~\cite{haroche1989cavity,walther2006cavity,berman1994cavity,haroche2013exploring}. Here, atoms can be trapped in the standing wave field of a lossy optical cavity (with atom-field coupling rate $g$ and photon loss rate $\kappa$) at equivalent positions, i.e., with equal couplings to the cavity light mode. The lossy cavity opens an additional decay channel along the cavity direction. This occurs in the limit of $\kappa\gg g$, where the excitation of the atoms is taken by the cavity field and emitted quickly through the side mirrors. This new decay channel is added to the intrinsic spontaneous emission into all three directions of free space and is characterized by a collective collapse operator $S$ (after adiabatic elimination of the cavity mode) and an associated collapse rate $\Gamma=g^2/\kappa$. Assuming that the cavity induced decay is much larger than the individual atom spontaneous emission, the dynamics of the atoms are fully described the master equation in Eq.~\eqref{master}. This is referred to as cavity-induced superradiance and has been implemented to realize a form of superradiant lasing~\cite{bohnet2012asteadystate,bohnet2014linear,norcia2016superradiance}. Moreover, waveguide and circuit quantum electrodynamics provide alternative, simpler to manipulate, platforms to realize Dicke superradiance. In these platforms, it is easier to design ensembles of practically non-interacting artificial electronic systems, placed at multiples of the wavelength of the resonator mode~\cite{Superradiance2016Lambert,Controllable2020Wang}. \par
\indent Mathematically, the Dicke superradiance problem involves determining the time evolution of the density operator $\rho(t)$
for $N$ two-level systems undergoing collective decay. The density operator follows the master equation in Lindblad form, presented here in a simplified version within the interaction picture, where the trivial effect of the free Hamiltonian has been eliminated:
\begin{equation}
\label{master}
    \dot{\rho}(t) = \mathcal{L}[\rho] = \Gamma\Bigg[ S \rho S^\dagger - \frac{1}{2}\Big(S^\dagger S \rho + \rho S^\dagger S \Big)\Bigg]=B[\rho]+C[\rho] .
\end{equation}
The separation of the Lindblad term into two superoperators is based on the observation that $B[\rho]=\Gamma S \rho S^\dagger$ lowers the excitation number by one, while $C[\rho]$ conserves the number of excitations in the system. The system of $N$ two-level system spans a Hilbert space of dimension $2^N$; the symmetric form of the collapse operator $S$, however, restricts the evolution solely within the symmetric subspace spanned by $N+1$ collective Dicke states $\ket{m}$, as the initially inverted state is part of this subspace. The index runs over $m=0,1,\ldots,N$. The ground $\ket{0}=\ket{gg...g}$ and fully excited state $\ket{N}=\ket{ee...e}$, are easily expressed in terms of the bare basis. The symmetric operators $S$ and $S^\dagger$ have the following action on the Dicke states
%%%%%%%%%%%%%%%%%%%%
\begin{subequations}
\begin{align}
    S | m \rangle &= \sqrt{h_m} |m-1 \rangle\\
    S^\dagger | m \rangle &= \sqrt{h_{m+1}} |m+1 \rangle,
\end{align}
\end{subequations}
%%%%%%%%%%%%%%%%%%%%%
keeping the evolution completely inside the symmetric subspace. The coefficients are defined as
\begin{equation}
h_m =m \bar m \qquad \text{with} \qquad \bar m=(N+1)-m
\end{equation}
and exhibit a double degeneracy for even $N$ as $h_m=h_{\bar m}$. For odd $N$, there is an additional non-degenerate point at $m=(N+1)/2$.\par

There are a few competing approaches to find the solution to the problem listed above. We present here five such approaches, and stress their advantages and shortcomings. For the first four approaches, we will recast the problem into its formal solution as an infinite time series expansion
\begin{equation} 
\label{expansion}
    \rho(t) = e^{\mathcal{L}t} [\rho_0] = \sum_{n=0}^\infty \frac{t^n}{n!} \mathcal{L}^n[\rho_0]=\sum_{n=0}^\infty \frac{t^n}{n!} \mathcal{L}[\cdots \mathcal{L}[\rho_0]].
\end{equation}
Thus, it is required to evaluate the repeated action of the superoperator $\mathcal{L}$ onto the initial density operator $\rho_0=\rho(0)=\ket{N}\bra{N}$. We remark that the density operator at any time is \textit{diagonal in the Dicke basis}: this can be seen by noticing the action of the two superoperators $B[\ket{m}\bra{m}]=\Gamma h_m \ket{m-1}\bra{m-1} $ and $C[\ket{m}\bra{m}]=-\Gamma h_m \ket{m}\bra{m}$, which indicates that no off-diagonal elements of the density operator can be generated during this incoherent evolution, when starting in a Dicke state. This allows to write $\rho(t) = \sum_{m=0}^N\rho_m(t) \ket{m}\bra{m}$, meaning that the task is reduced to finding the expression of $\rho_m(t)$ for any time.\par

Four of the approaches introduced in the following involve the time expansion of the density operator as in Eq.~\eqref{expansion} and are based on:
\begin{enumerate}
    \item \textbf{Solving a set of recursive equations}: We find a recursive equation of the form
\begin{equation}
    \rho_{m-1}^{(n+1)} = -h_{m-1} \rho^{(n)}_{m-1} + h_{m} \rho^{(n)}_{m},
\end{equation}
where we defined 
\begin{equation} \label{power-series}
    \rho_m(t) = \langle m | e^{\mathcal{L}t} \rho_0|m\rangle =  \sum_{n=0}^\infty \frac{(\Gamma t)^n}{n!} \rho_m^{(n)}.
\end{equation}
    The recursion connects the $n+1$ component in the time series of state $m-1$ to the $n$ components stemming either "longitudinally" from the same state $m-1$ or "diagonally" from the state situated immediately above $m$. We then construct a solution $\rho^{(n)}_m$ for any order $n$ and any state $m$ and sum it from zero to infinity, to construct the density operator as an infinite time series.
    \item \textbf{Combinatorics approach}: We formally split the time series into a binomial expansion of non-commuting superoperators
    \begin{equation} 
    \rho(t) =  \sum_{n=0}^\infty \frac{t^n}{n!} \Big(B+C\Big)^n[\rho_0].
    \end{equation}
   Writing out the contributions of each term in the expansion leads to a sum over all possible paths, which we then add to find the exact expression for any $\rho_m(t)$.
    \item \textbf{Probabilistic approach}: We expand $e^{\mathcal{L}\Delta t}$ for small $\Delta t$ in order to connect the density operator between discretized times $j$ and $j+1$, with $t=j\Delta t$ to $t+\Delta t=(j+1)\Delta t$. Two complementary events emerge, one quantified by $d_j$, which is the probability to perform a jump within the time interval $\Delta t$ and $s_j=1-d_j$, quantifying the probability to remain in the respective state. Summing up all possible paths leading from the initially inverted state $\rho_0$ to a target state $m$ leads to an analytical expression for $\rho_m(t)$. 
    \item \textbf{Non-Hermitian Hamiltonian approach}: We start again with the recursive equations and cast it into a form involving a non-Hermitian Hamiltonian~$H$. With this, the general solution involves a matrix exponential requiring a Jordan decomposition. Due to the symmetry $h_m = h_{\bar{m}}$, $H$~exhibits non-Hermitian degeneracies called exceptional points, which result in a modified decay compared to a non-degenerate system. 
\end{enumerate}
\indent The approaches described above allow for the construction of the exact solution $\rho(t)$ with different degrees of difficulty and ad-hoc observations involved in the derivations. The initial method considered in Ref.~\cite{holzinger2024exact}, in particular involved an ad hoc step to allow the deduction of the general solution from solving the rate equations for states close to $N$. However, this approach, while cumbersome to derive and follow, indicated that the time evolution of the density operator can be written in an extremely compact form as the residue sum of a given function. While some approaches above can be tailored to result in residue sums and integration in the complex plane via generating functions or (inverse) Laplace transforms, we will focus in detail on a technique involving an unraveling of the master equation into quantum trajectories. This leads easily to a solution for any $\rho_m(t)$ as a closed contour integral in the complex plane.

\begin{enumerate}
    \setcounter{enumi}{4}
    \item \textbf{Quantum jump approach}: We proceed with a quantum jump unraveling of the master equation in Eq.~\eqref{master}. Each trajectory is described by a state vector $\ket{\psi(t)}_j$, where the $j$ trajectory presents $N$ jumps at times $t_N,t_{N-1},\ldots,t_1$ describing the system's evolution from the initial state $\ket{N}$, all the way to the ground state $\ket{0}$. The times of the jumps are dictated by the choice of an $N$-tuple of randomly generated numbers from a uniform distribution on the $[0,1]$ interval. An average over an infinite number of trajectories leads to an exact solution of the master equation as a closed contour integral.
\end{enumerate}

%%%%%%%%%%%%%%%%
\begin{figure}
\centering
\includegraphics[width=0.85\columnwidth]{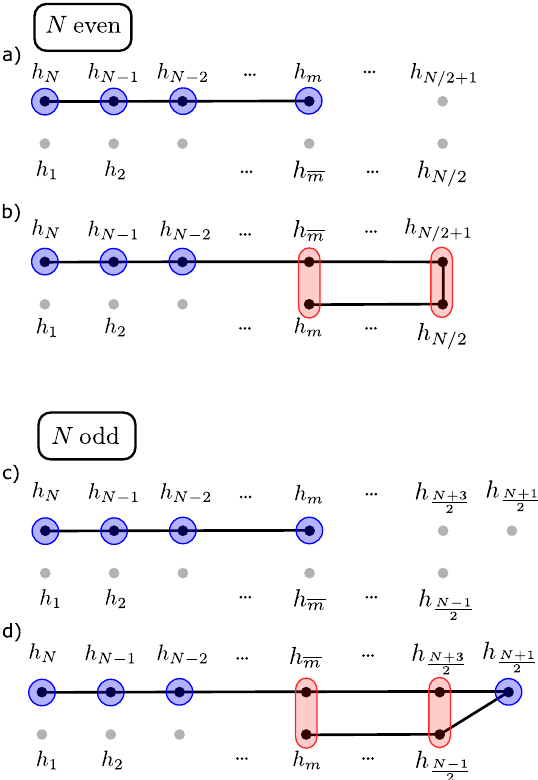}
    \caption{Procedure for adding up residues to form the solution of the density matrix element $\rho_m$ in Eq.\eqref{eq:rho_m residues}. The possible poles are ordered from left to right according to their magnitude. Blue marks single poles, red marks double poles, and gray poles which are not included in the solution of a given state $m$. a) For even $N$ and $m \geq (N+1)/2$ only single poles contribute to the solution $\rho_m(t)$, which is marked by the solid line starting at state $h_m$ and going to state $N$. b) For $\bar{m} \geq (N+1)/2$ some of the poles are double, shown as pairs in red. c) and d) For $N$ is odd, the sole difference is that the pole at $h_{(N+1)/2}$ is always a simple pole.}
    \label{fig:residues}
\end{figure}
%%%%%%%%%%%%%%%%%%
%%%%%%%%%%%%%%%%%%%%%%
\section{The analytical solution}
\label{Solution}
%%%%%%%%%%%%%%%%%%%%%%
Before diving into the details of the analytical derivations proposed here, let us first express the solution to Dicke superradiance in a compact form, as it naturally comes out from the quantum jump approach. Noting that the density operator during the evolution is always in diagonal form, as no off-diagonal elements can be created, $\rho(t)=\sum_{m=0}^N \rho_m(t)\ket{m}\bra{m}$, we can write the solution for the time evolution of any diagonal element as
\begin{equation}
\rho_m(t)=\frac{1}{2\pi\iu}\oint_{\mathcal C} \mathrm{d}z \,f_m(z,t)=\!\!\!\sum_{p\in\text{poles}} \!\!\!\text{Res}\Big[f_m(z,t)\Big]_{z=p},
\label{eq:rho_m residues}
\end{equation}
where the sum of residues runs over all poles $p$, simple or double depending on the state $m$. This is a closed contour integral in the complex plane of the function
\begin{equation}
\label{Function}
f_m(z,t)=(-1)^{N-m}\frac{h_N\cdots h_{m+1}}{(z-h_N)\cdots(z-h_m)}\me^{-z\Gamma t}.    
\end{equation}
For a given state $m$, the contour integral is evaluated as shown in Fig.~\ref{fig:residues}, where the cases $N$ odd and even are distinguished from each other. To obtain the time evolution of a state $m$ state above the equator, the poles to be summed over run from $h_m$ to $h_N$ and are simple poles (color coded in blue) obeying the formula 
\begin{equation}
\text{Res}\Big[f_m(z,t)\Big]_{z=h_p} = \lim_{z \to h_p}\Big[f_m(z,t)\left(z - h_p\right)\Big].
\end{equation} 
For states below the equator, contributions from any degenerate pairs $h_m$ and $h_{\bar{m}}$ (color coded in red) are only added once, as double poles are computed using the following formula for the residue:
\begin{equation}
\text{Res}\Big[f_m(z,t)\Big]_{z=h_p} = \lim_{z \to h_p}\dv{z}\Big[f_m(z,t)\left(z - h_p\right)^2\Big].
\end{equation}
\indent Let us remark that this solution admits a straightforward generalization to the time evolution of a system obeying Eq.~\eqref{master} and initialized in any of the Dicke states. For an initial state $\rho_0=\ket{m_0}\bra{m_0}$, with $1\leq m_0 \leq N$, the solution reads
\begin{equation} \label{any-state}
    \rho_m(t) = \sum_{p\in\text{poles}} \text{Res}\Big[f_{m_0,m}(z,t)\Big]_{z=p}.
\end{equation}
Notice that the sum over the residues runs only between the target state $m$ and $m_0$ and the function is redefined as
\begin{equation}
\label{eq:fm0m}
f_{m_0,m}(z,t) = (-1)^{m_0-m}\frac{ h_{m_0}\cdots h_{m+1}}{(z-h_{m_0})\cdots(z-h_m)}e^{-z\Gamma t}.
\end{equation}
This result allows an analytical description of the time evolution of the full density matrix for any initial mixture of Dicke states.

\section{Quantum trajectories approach} 
The master equation in Eq.~\eqref{master} can be unraveled in terms of quantum trajectories for the state vector of the system subjected to non-Hermitian dynamics and discrete or continuous state jumps~\cite{breuer2002theory,walls2006quantumoptics}. We choose here the quantum jump approach, where the state vector $\ket{\psi(t)}_j$ for any unraveling $j$ undergoes quantum jumps, under the action of the collapse operator $S$ and at rate $\Gamma$. The validity of this unraveling is guaranteed by the fact that the  exact time evolution of the density operator is recovered as the average
\begin{equation}
    \rho(t)=\frac{1}{n_\text{traj}}\sum_{j=1}^{n_\text{traj}} \ket{\psi(t)}_j\bra{\psi(t)}_j,
\end{equation}
over an infinite number of trajectories $n_\text{traj}$. Typically, this approach is purely numerical as it sometimes allows for computational speed-ups by considering a reduced set of trajectories (presenting a clear advantage to the full master equation simulation).\par
\begin{figure*}[t]
    \centering   
\includegraphics[width=1.95\columnwidth]{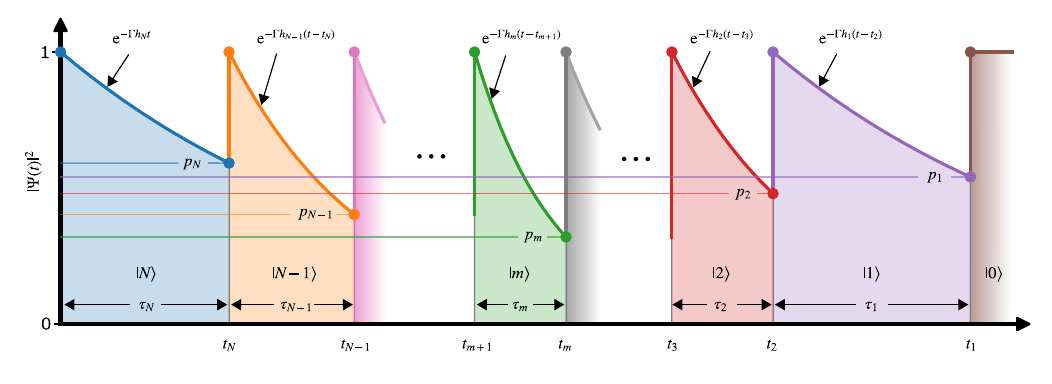}
\caption{Time evolution of the system during a given trajectory characterized by a set of $N$ jumps occurring at times $t_N$,...,$t_1$ taking the system from the initial state $\ket{N}$ at time $t=0$ to the ground state $\ket{0}$. The time intervals between two consecutive jumps are denoted by $\tau_N$,...,$\tau_1$ and are computed from the requirement that $e^{-\Gamma h_{m} \tau_{m}}=p_{m}$, for any state $m$. The population of state $m$, as encoded in $\rho_m(t)$ is estimated by averaging the time the quantum system spends on state $\ket{m}$ over all possible trajectories.}
    \label{fig2}
\end{figure*}

\indent Let us describe the standard procedure for unraveling the master equation Eq.~\eqref{master} in the quantum jump approach. A particular trajectory is generated by first picking a set of mutually independent random numbers between $0$ and $1$, arranged as $p_N,p_{N-1},...,p_1$. A given $p_m$ value characterizes the probability for the occurrence of a jump from state $m$ to $m-1$. Each trajectory contains $N$ jumps taking the system from the fully inverted state $\ket{N}$ to the ground state $\ket{0}$ (see Fig.~\ref{fig2} for illustration). The full procedure for generating a given trajectory is then
\begin{itemize}
    \item pick a random number $p_N$ uniformly distributed between 0 and 1
    \item evolve the state vector from $\ket{\psi(0)}=\ket{N}$ for time $t_N=\tau_N$ with the non-Hermitian Hamiltonian $H_\text{nH}=-i\frac{\Gamma}{2} S^\dagger S$ ($\hbar=1$) 
    \item perform the first jump at time $t_N=\tau_N$, computed from the requirement that the norm of $\ket{\psi(\tau_N)}$ reaches the value set by $p_N$, i.e. $\braket{\psi(\tau_N)|\psi(\tau_N)}=e^{-\Gamma h_N \tau_N}=p_N$
    \item renormalize the state vector to unity and proceed again starting from the time $t_N$ to the time $t_{N-1}=t_N+\tau_{N-1}$ by again checking that the norm has reached the randomly picked value $p_{N-1}$, i.e. asking that $e^{-\Gamma h_{N-1} \tau_{N-1}}=p_{N-1}$ 
    \item repeat the procedure until the ground state $\ket{0}$ is reached at time $t_1$, with the last time interval derived from $e^{-\Gamma h_1 \tau_1}=p_1$.
\end{itemize}
Each trajectory is fully determined by the values in the $N$-tuple $(\tau_N,...,\tau_1)$, which are determined by the random choice of the generated $N$-tuple $(p_N,...,p_1)$. Notice that the average over an infinite number of trajectories, which is needed to construct the exact expression for the evolution of the density operator, is equivalent to an integral over all possible values of the $p_m$'s within $0$ and $1$ with a flat probability distribution (since the numbers are picked randomly in this interval). 

%%%%%%%%%%%%%%%%%%%%%%%%%%%%%%%%%%%%%%%%%%%%%%%%%%%%%%%%%%%%%%%%%%%%%
%%%%%%%%%%%%%%%%%%%%%%%%%%%%%%%%%%%%%%%%%%%%%%%%%%%%%%%%%%%%%%%%%%%%%
For a given trajectory, the system spends a total time $\tau_{m}$ in state $\ket{m}$, with $t$ enclosed between $t_{m+1}$ (the time at which the jump from state $m+1$ into the state $m$ occurs) and $t_m$ (when the next jump out of the state occurs). An average of all possible trajectories then gives the value of $\rho_m(t)$ as the probability that the system finds itself in state $m$ at time $t$. This corresponds to an average of the segment length $\left[\theta(t-t_{m+1})-\theta(t-t_{m})\right]$ (where $\theta$ denotes the Heaviside function) over all possible values of $p_N$,...,$p_1$. The evolution is visually represented in Fig.~\ref{fig2}. The population of the state $\rho_m(t)$ can then be computed as
\begin{equation}
    \rho_m(t)=\int_0^1 \text{d}p_{N}\cdots\int_0^1 \text{d}p_{m}\left[\theta(t-t_{m+1})-\theta(t-t_{m})\right],
\end{equation}
where we have used the fact that the probability distribution of any of the random variables is flat within the interval $[0,1]$. Notice that, as expected, when summing the probability of being in any of the segments $\tau_m$ with $m=0,...,N$, one obtains unity, as the trace of the density operator is preserved $\sum_{m=0}^N\rho_m(t)=1$. We proceed with a change of integration variables from $p_{k}$ to $\tau_k$, such that $\mathrm{d}p_k=-\Gamma h_k e^{-\Gamma h_k\tau_k}\mathrm{d}\tau_k$, for any $k=1,..N$. With the corresponding change in the integration limits for $\tau_k$, we can write
\begin{widetext}
\begin{equation}
\label{long}
\rho_m(t)=\Gamma^{N-m+1}h_N \cdots h_m\int_0^\infty \text{d}\tau_{N}e^{-\Gamma h_N\tau_N} \cdots\int_0^\infty \text{d}\tau_{m} e^{-\Gamma h_m\tau_m} \left[\theta\Big(t-\sum_{k=m+1}^N\tau_k\Big)-\theta\Big(t-\sum_{k=m}^N\tau_k\Big)\right],
\end{equation}
\end{widetext}
where we replaced $t_{m}=\sum_{k=m}^N\tau_k$ and similarly for $t_{m+1}$. Let us split the integral above into two parts, by writing $\rho_m=\mathcal{I}_{m+1}-\mathcal{I}_m$.
We continue by remarking that the Heaviside theta function can be expressed in integral form as follows 
\begin{equation}
    \theta(x)=\frac{1}{2\pi\iu}\lim_{\epsilon\to 0^+}\int_{-\infty}^{\infty}\text{d}s\frac{\me^{\iu \Gamma s x}}{s-\iu\epsilon}.
\end{equation}
We substitute this into Eq.~\eqref{long} with the replacement $x$ by $t-\sum_{k=m+1}^N\tau_k$ first, to evaluate $\mathcal{I}_{m+1}$ and then with the replacement $x$ by $t-\sum_{k=m}^N\tau_k$ to evaluate $\mathcal{I}_{m}$. We distribute the terms with $\tau_k$ from the exponent into all integrals over $\tau_k$. This allows to compute any of the integrals which lead to the generation of poles of $s$ at \mbox{$-\iu h_k$}:
\begin{equation}
    \int_0^{\infty}\text{d}\tau_{k} e^{-\Gamma h_k\tau_k}e^{-\iu s\Gamma \tau_k}=\frac{1}{(\iu s+h_k)\Gamma}.
\end{equation}
The multiplication of $N-m+1$ of such integrals leads to a term proportional to $\Gamma^{N-m+1}$ in the denominator, which cancels the same term in the numerator. Substituting this result into Eq.~\eqref{long} allows to derive:

\begin{equation}
\mathcal{I}_{m+1}=\frac{1}{2\pi \iu}\lim_{\epsilon\to 0^+}\int_{-\infty}^{\infty} \text{d}s\left[\prod_{k=m+1}^{N}\frac{h_k}{\iu s+h_k}\right]\frac{\me^{\iu t \Gamma s}}{s-\iu\epsilon},
\end{equation}
and 
\begin{equation}
\mathcal{I}_{m}=\frac{1}{2\pi \iu}\lim_{\epsilon\to 0^+}\int_{-\infty}^{\infty}\text{d}s\left[\prod_{k=m}^{N}\frac{h_k}{\iu s+h_k}\right]\frac{\me^{\iu t \Gamma s}}{s-\iu\epsilon}.
\end{equation}
Both integrals can be extended into the complex plane with the observation that all poles are in the lower plane. This means that we can already take the limit $\epsilon\rightarrow 0$ with no effect. Next, we change variables to $z=-\iu s$ and take the difference of the integrals above to lead to the solution
%%%%%%%%%%%%%%%%%%%%%
\begin{equation}
\rho_m=\frac{1}{2\pi\iu}\oint_{\mathcal C} \mathrm{d}z \, f_m(z,t),
\end{equation}
%%%%%%%%%%%%%%%%%
where the integral is over a contour including all poles of the following function
\begin{equation}
f_m(z,t)=(-1)^{N-m}\frac{h_N\cdots h_{m+1}}{(z-h_N)\cdots(z-h_m)}\me^{-z\Gamma t}.  
\end{equation}
This function is holomorphic everywhere in the complex plane, which means that the integral is given by the residue sum.
%%%%%%%%%%%%%%%%%%%%%%%%%%%%%%%%%%
%%%%%%%%%%%%%%%%%%%%%%%%%%%%%%%%%%%%
\section{A combinatorics approach} 
%%%%%%%%%%%%%%%%%%%%%%%%%%%%%%%%%%%%%%
%%%%%%%%%%%%%%%%%%%%%%%%%%%%%%%%%%%%%%
Let us now return to the time series expansion, where we separate $\mathcal{L}[\rho]$ into two superoperators, either conserving the excitation number ($C[\rho]$) or lowering it ($B[\rho]$), such that the density operator at any time is

\begin{equation} 
\mathcal{L}^n[\rho_0]=\sum_{r=0}^n \bigg[ \sum_{\substack{j_N+\ldots+j_{N-r}\\=n-r}}\!\!\!\!\!\!\!\!\!\overbrace{C .. C}^{j_{N-r}}B \ldots B\overbrace{C .. C}^{j_{N-1}}B\overbrace{C .. C}^{j_N}[\rho_0]\bigg].
\end{equation}

In order to obtain the most general form of the sequences above, we first placed $r$ superoperators $B$ in the sequence (with $r$ varying from $0$ to $n$) from right to left and added paddings of $C$s in the spaces between the $B$s, with $j_N$, $j_{N-1}$,\ldots, $j_{N-r}$ counting the numbers of superoperators $C$ in the sequence. Summing over $r$ from $0$ to $n$ and for any possible sequences (or spellings) corresponding to each fixed $r$ gives all possible combinations involved in the binomial expansion.\par
\indent First, we eliminate the $C$ paddings. We notice that 
\begin{equation}
  C^{j_N}[\rho_0]=(-\Gamma h_N)^{j_N}\rho_0.  
\end{equation}
 After the action of the lowering superoperator, the paddings of $C$s, simply have the same effect, but on a lower state:
\begin{equation}
 C^{j_{N-1}}B[\rho_0]=(-\Gamma h_{N-1})^{j_{N-1}}B[\rho_0]
\end{equation}
This is then extended for any state, all the way to the ground state. We can start by applying the $C$ operators starting from the right side and then collect all the terms and compress the resulting $B$ terms to the right. Collecting all terms and with $(-\Gamma)^{j_N+\ldots+j_{N-r}}=(-\Gamma)^{n-r}$ leads to the following result

\begin{equation} 
\mathcal{L}^n[\rho_0]=\sum_{r=0}^n \bigg[ (-\Gamma)^{n-r}B^r[\rho_0] \! \! \! \! \! \! \sum_{j_N+\ldots+j_{N-r}=n-r} \! \! \! \! \! \! h_N^{j_N}\cdots h_{N-r}^{j_{N-r}}\bigg].
\end{equation}

In order to evaluate the time evolution of the matrix element $\rho_m(t)$ we will sandwich the expression above with $\ket{m}$ and $\bra{m}$ and use the fact 
\begin{equation}
    \bra{m}B^r[\rho_0]\ket{m}=h_N \cdots h_{m+1}\delta_{r,N-m},
\end{equation}
where $\delta_{m,m'}$ is the Kronecker delta. Summing up the infinite time series implies the need for performing the constrained sum:
\begin{equation} 
\sum_{j_N+\ldots+j_{N-r}=n-r}h_N^{j_N}\cdots h_{N-r}^{j_{N-r}}.
\end{equation}
These kind of sums can be obtained from the expansion of given generating functions, as we describe in App.~\ref{App.:A}. We make use of the result obtained there, with the identifications $n_t=N-m+1$ and $M=n-(N-m)$ and identifying the following generating function
\begin{equation}
f_{N-m+1}(z)=z^{N-m+1}\bigg(\frac{1}{z-h_N}\bigg)\cdots\bigg(\frac{1}{z-h_m}\bigg).
\end{equation}
The constrained sum is the coefficient of the $1/z^{n-(N-m)}$ term in the series expansion of the function $f_{N-m+1}(z)$ and can be computed as the sum of the residues of $f_{N-m+1}(z)e^{z \Gamma t}$. This gives then evidently the result listed in Eq.~\eqref{eq:rho_m residues}.

%%%%%%%%%%%%%%%%%
%%%%%%%%%%%%%%%%%
\section {A probabilistic approach} 
%%%%%%%%%%%%%%%%%
%%%%%%%%%%%%%%%%%
To compute the density matrix element $\rho_m(t)$, we discretize $t=j\Delta t$. During each very small interval $\Delta t$, the system can either jump to the lower level or stay on the same initial level. This results from the short time expansion of $e^{\mathcal{L} \Delta t}\sim 1+\mathcal{L} \Delta t$, which applied to state $\ket{k}$ leads to
\begin{equation}
    e^{\mathcal{L} \Delta t}\ket{k}\bra{k}\sim (1-h_k\Gamma \Delta t) |k\rangle \langle k| + h_k\Gamma \Delta t |k-1\rangle \langle k-1|.
\end{equation}  
This indicates a probability $d_k=h_k \Gamma \Delta t$ for the system to perform a jump during the time interval $\Delta t$, when on a given state $\ket{k}$. We denote the complementary probability, of staying on the level, with $s_k=1-h_k\Gamma \Delta t$. Let us now estimate the probability for the system to be in state $\ket{m}$ at time $t$, i.e. after $j$ steps. We note that this requires a number of $N-m$ jumps, with associated probabilities from $d_N$ to $d_{m+1}$. We can then write any sequence which contributes to the population of the state $m$ as
\begin{equation}
\! \underbrace{s_N\cdots s_N}_{j_N}d_N\underbrace{s_{N-1} \cdots s_{N-1}}_{j_{N-1}}d_{N-1} \cdots d_{m+1}\underbrace{s_m\cdots s_m}_{j_m}
\end{equation}
and perform the sum over all sequences

\begin{equation}
\rho_m(t)=\! \! \! \! \! \! \! \! \! \! \sum_{j_N+\ldots+j_m+(N-m)=j} \! \! \! \! \! \! \! \! \! \! s_N^{j_N} d_N s_{N-1}^ {j_{N-1}}d_{N-1}\cdots d_{m+1}s_m^{j_m}.
\end{equation}

The constraint $j_N+\ldots+j_m+(N-m)=j$ ensures that the final time is $t=j\Delta t$ and that $(N-m)$ jumps have occurred in this time interval, leading to the target state $\ket{m}$. The numbers $j_N,\ldots,j_m$ count the number of time steps the system stays in the corresponding state between jumps. Notice that the product over jumps leads to $h_N\cdots h_{m+1}\Gamma^{N-m}(\Delta t)^{N-m}$ and we can write 
%%%%%%%%%%%%%%%%%%%%%%%
\begin{equation}
\rho_m(t)=\mathcal{U}_m \Gamma^{N-m}(\Delta t)^{N-m} \!\!\!\!\!\!\!\!\! \sum_{j_N+\ldots+j_m+(N-m)=j} \!\!\!\!\!\!\!\!\!\!\!\!\!s_N^{j_N}\cdots s_m^{j_m},
\end{equation}
%%%%%%%%%%%%%%%%%%%%%%
where we denoted $\mathcal{U}_m=h_N\cdots h_{m+1}$.
In order to add all possible paths, we make use of the generating function from App.~\ref{App.:A} with the replacements $n_t \rightarrow N-m+1$ (number of terms in the sum), $M\rightarrow j-(N-m)$ and $a_1,\ldots,a_{n_t}$ replaced by $s_N,\ldots,s_m$. The solution can then be written as
%%%%%%%%%%%%%%%
\begin{equation}
    \rho_m(t)=
\sum_{p \in \text{poles}} \text{Res}\Big[z^j\Big(\frac{h_N\Gamma \Delta t}{z-s_N}\Big)\cdots \Big(\frac{h_m\Gamma \Delta t}{z-s_m}\Big) \Big]_{z=p}.
\end{equation}
%%%%%%%%%%%%%%%%%
Notice that when evaluating $z$ at the poles, the differences in the denominator are $s_k-s_{k'}=-\Gamma\Delta t(h_k-h_{k'})$ canceling the terms containing $\Gamma\Delta t$ in the numerator and also leading to a term $(-1)^{N-m}$. In the limit of infinitesimal time steps, $\Delta t \to 0$, $j=t/\Delta t$ can be taken to infinity. In this limit, any term $s_k^j=(1-h_k\Gamma t/j)^{j}$ can be approximated by its limit value $e^{-\Gamma h_k t}$. The expression above then again maps fully into the solution presented in Sec.~\ref{Solution}.
%%%%%%%%%%
\section{Rate equations approach} 
\label{sec:rate-equations-aproach}
%%%%%%%%%%%%%%%%%%%%%%%%%%%%%%%%%%
We start by noticing the following rule for the application of the Lindblad superoperator onto a given Dicke state
\begin{equation}
    \mathcal{L}[|m\rangle \langle m|] = -h_m |m\rangle \langle m| + h_m |m-1\rangle \langle m-1|,
\end{equation}  
indicating that the collective decay moves the system either horizontally within the same Dicke state at a rate $h_m$ or diagonally to the lower Dicke state with $m-1$ at a rate $h_m$. Of course, at $m=0$, at the bottom of the ladder, we have $\mathcal{L}[|0\rangle \langle 0|]=0$. We can then find a general recursion rule
\begin{equation}
    \label{eq:rate-equation}
    \rho_{m-1}^{(j+1)} = -h_{m-1} \rho^{(j)}_{m-1} + h_{m} \rho^{(j)}_{m},
\end{equation}
which shows that any order $j+1$ in the time power series can be computed from the previous order $j$ with knowledge of all the terms $ \rho_{m}^{(j)}$. Starting with all emitters excited at time $t=0$ means that $\rho_N^{(0)}=1$, which allows one to compute $\rho_N^{(1)}=-h_N$ and so on, generally leading to $\rho_N^{(j)}=(-h_N)^j$. The infinite sum leads to the expected result $\rho_N(t) =\mathrm{exp}(-h_N \Gamma t)= \mathrm{exp}(-N \Gamma t)$.\par
\indent Solving the recursive equation involves guessing the general expression by extending the form of the solution for $N-1$, $N-2$ and so on (for a fully detailed discussion see Ref.~\cite{holzinger2024exact}). This expression is the following
\begin{equation}
\label{SolPlus}
    \rho_{m}(t) =  \sum_{j=m}^N (-1)^{N-m} h_N\cdots h_{m+1}\Bigg[\prod_{\substack{j'=m \\j'\neq j}}^N \frac{1}{h_j-h_{j'}}\Bigg] e^{-h_j\Gamma t},
\end{equation}
and proves correct for any point $m\geq N/2+1$ (for even $N$) but it fails below the equator. However, as its form indicates that the results is a sum over residues, the extension to the form introduced in Sec.~\ref{Solution} proves to give the right solution for any state during the time evolution.

%%%%%%%%%%%%%%%%%
%%%%%%%%%%%%%%%%%
\section{Non-Hermitian Hamiltonian approaches} 
%%%%%%%%%%%%%%%%%%
%%%%%%%%%%%%%%%%%%
The recursion equation in Eq.~\eqref{eq:rate-equation} can be turned into a set of coupled rate equations which will see the master equation in Eq.~\eqref{master} cast into the following matrix form
\begin{equation}
    \label{eq:diff}
    \dot{\boldsymbol{\rho}}(t) = \Gamma H \cdot \boldsymbol{\rho}(t).
\end{equation}
The vectorized density matrix is $\boldsymbol{\rho}(t) = (\rho_N(t),\rho_{N-1}(t),$ $\ldots,\rho_1(t),\rho_0(t))^T$ (the superscript~$T$ denotes the transposition). The initial condition is $\boldsymbol{\rho}(0)=\boldsymbol{\rho}^{(0)}$. The matrix $H$ is given by
\begin{equation}
    H =
    \left(
\begin{array}{rrrrrr}
 -h_N & & & & & \\
 h_N & -h_{N-1} & & & & \\
 & h_{N-1} & \ddots & & & \\
 & & \ddots & -h_2 & & \\
 & & & h_2 & -h_1 & \\
 & & & & h_1 & -h_0 \\
\end{array}
\right),
\end{equation}
which solely depends on the coefficients $h_j$ (with the reminder that $h_j=h_{\bar{j}}$ and $h_0=0$). It is clear that $H$ is non-Hermitian ($H \neq H^\dagger$), and that the spectrum can be read off the diagonal of $H$.
\subsection{Time evolution operator}
The equation above Eq.~\eqref{eq:diff} admits the formal solution $\boldsymbol{\rho}(t) = U(t) \cdot \boldsymbol{\rho}^{(0)}$, with the time evolution operator $U(t)=e^{H \Gamma t}$. 
To compute the matrix exponential, one has to determine the Jordan normal form $J$ of $H$ via
\begin{equation}
    \label{eq:jordandecomp}
    H = T \cdot J \cdot T^{-1},
\end{equation}
where $T$ is the similarity matrix transforming between $J$ and $H$. In general, $J$ is a direct sum of Jordan blocks; in our case, due to the at most two-fold degenerate eigenvalues of $H$, the Jordan blocks are at most of size two. In App.~\ref{App.:B}, we give details on the determination of $T$ and $T^{-1}$, as well as show that all Jordan blocks in $J$ associated with a two-fold degenerate $-h_j$ are of size two. \par
In the recent non-Hermitian physics literature (for a review, see Ref.~\cite{RevModPhys.93.015005}), a point in parameter space where the Jordan normal form $J$ contains a Jordan block of size $n$ is referred to as an exceptional point (EP) of order~$n$ (EP$n$). In this language, our system exhibits $\lfloor N/2\rfloor$ (floor of $N/2$) EP$2$s.\par
The explicit forms of the Jordan blocks of size one and two are, respectively,
\begin{equation}
    J_1(\lambda) = \begin{pmatrix}
        \lambda
    \end{pmatrix},
    \qquad J_2(\lambda) =
    \begin{pmatrix}
        \lambda & 1 \\
        0 & \lambda
    \end{pmatrix},
\end{equation}
and $J$ is for even $N$ (see App.~\ref{App.:B} for odd $N$) by
\begin{equation}
    \label{eq:Jeven}
    J = J_2 (-h_{N/2+1}) \oplus \cdots \oplus J_2(-h_N) \oplus J_1 (0).
\end{equation}
Then, the time evolution operator is given by
\begin{equation}
    \label{eq:Ut}
    U(t) = T \cdot e^{J \Gamma t} \cdot T^{-1},
\end{equation}
where now $e^{J \Gamma t}$ is now a direct sum of the blocks $e^{J_1(-h_j)\Gamma t} = (e^{-h_j \Gamma t})$, and
\begin{equation}
    e^{J_2(-h_j)\Gamma t} =
    \begin{pmatrix}
        e^{-h_j \Gamma t} & \Gamma t \,e^{-h_j \Gamma t} \\
        0 & e^{-h_j \Gamma t}
    \end{pmatrix},
\end{equation}
where we see that the top right element is not a pure exponential decay. These decays of the form $\Gamma t e^{-h_m\Gamma t}$ stem from the occurring EP$2$s.
\subsection{Laplace transform}
Another way to solve Eq.~\eqref{eq:diff} is by going to Laplace space. Laplace transforming both sides of Eq.~\eqref{eq:diff} with respect to $\Gamma t$ for a single element of the density operator $\rho_m(t)$ (with Laplace transform $\bar \rho_m (z)$) leads to
\begin{equation}
    z \bar\rho_m(z) - \rho_m^{(0)} = \sum_{m'} H_{mm'}\bar\rho_{m'}(z),
\end{equation}
which can be rewritten as
\begin{equation}
    \sum_{m'}\left( z \delta_{mm'} - H_{mm'} \right)\bar \rho_{m'}(z) = \rho_m^{(0)}.
\end{equation}
This can be cast for all $m$ in the vector form $(z\mathbb{1} - H)\cdot \boldsymbol{\bar\rho}(z) = \boldsymbol{\rho}^{(0)}$
where $\mathbb{1}$ is the $(N+1) \times (N+1)$ identity matrix. This can be solved in Laplace space by matrix inversion as
\begin{equation}
    \label{eq:solution-laplace}
    \boldsymbol{\bar\rho}(z)= \left(z\mathbb{1}-H \right)^{-1}\cdot\boldsymbol{\rho}^{(0)},
\end{equation}
where $\left(z\mathbb{1}-H \right)^{-1}\equiv R(z)$ is the so-called resolvent of $H$~\cite{dunford_linear_1988}. We can generally find all the matrix elements $R_{mm'}(z)$ of the resolvent (see App. \ref{App.:C} for details). However, here we restrict the discussion to the initially fully inverted case, i.e., $\rho_{m'}^{(0)}=\delta_{m'N}$, which means that the Laplace transform element $m$ is given by the simplified expression
\begin{equation}
    \label{eq:resolvent}
   \bar\rho_m(z)= R_{mN}(z) = \frac{1}{z+h_m}\prod_{j=m+1}^{N} \frac{h_j}{z+h_j}.
\end{equation}
To go back to the time domain, one has to take the inverse Laplace transform, which yields
\begin{align}
    \rho_m(t) = \frac{1}{2 \pi \mathrm{i}}\oint_{\mathcal C} \mathrm{d}z \, R_{mN}(z) e^{z\Gamma t},
\end{align}
where the contour $\mathcal C$ encloses all singularities of $R_{mN}(z)$.

%%%%%%%%%%%%%%%%%%%%%%%%%%
%%%%%%%%%%%%%%%%%%%%%%%%%%
\section {Conclusions and outlook} 
%%%%%%%%%%%%%%%%%%%%%%%%%%
%%%%%%%%%%%%%%%%%%%%%%%%%%
We have provided a number of different approaches leading to the same analytical solution for Dicke superradiance, namely the time evolution of a fully inverted ensemble under collective decay. The solution can also be extended to consider any initial state on the surface of the Bloch sphere as discussed in more detail in Ref.~\cite{holzinger2024exact}. As the permutational symmetry renders the problem simple enough, given that the dynamics is restricted to the symmetric subspace, it is not surprising that the problem admits a compact analytical solution. The quantum jump approach in particular gives an extremely elegant and simple way to obtain the solution as a contour integral in the complex plane.\par
\indent We expect, that the existence of analytical solutions for the time evolution of driven, dissipative open quantum systems is not restricted to this particular problem. The approaches proposed here might find applicability to more complex problems such as, among others: i) the coherently driven Dicke superradiant system, or ii) the incoherently and independently pumped Dicke superradiant system. In both cases, off-diagonal elements of the density operator are generated rendering the problem more complex. The second case is a limiting case of lasing, where the optical resonator can be eliminated owing to its large decay rate and the gain medium behaves superradiantly, thus describing a superradiant laser with the coherence stored in the gain medium instead of the cavity field~\cite{bohnet2012asteadystate,bohnet2014linear,norcia2016superradiance}. \\

\noindent \emph{Acknowledgments -} We acknowledge financial support from the Max Planck Society and the Deutsche Forschungsgemeinschaft (DFG, German Research Foundation) -- Project-ID 429529648 -- TRR 306 \mbox{QuCoLiMa}
(``Quantum Cooperativity of Light and Matter''). R.H. acknowledges funding by the Austrian Science Fund (FWF) 10.55776/W1259. J.T.G. acknowledges funding from the Max Planck Society Lise Meitner Excellence Program~\mbox{2.0}.

%TC:ignore

%%%%%%%%%%-----------%%%%%%%%%%-----------%%%%%%%%%%-----------%%%%%%%%%% Supplementary Material

\bibliography{apssamp}

\appendix
\clearpage
\pagebreak
\onecolumngrid

\section{Summation formulas}
\label{App.:A}
Let us consider the following generating function containing a product of $n_t$ (number of terms) functions and write its series expansion using the geometric series formula:
\begin{equation}
f_{n_t}(z)=z^{n_t}\bigg(\frac{1}{z-a_1}\bigg)\cdots\bigg(\frac{1}{z-a_{n_t}}\bigg)=\sum_{i_1=0}^{\infty}\ldots\sum_{i_{n_t}=0}^{\infty}a_1^{i_1}\cdots a_{n_t}^{i_{n_t}}\frac{1}{z^{i_1+\ldots+i_{n_t}}}=\sum_{n=0}^{\infty}\Bigg[\sum_{i_1+\ldots +i_{n_t}=M}a_1^{i_1}\cdots a_{n_t}^{i_{n_t}}\Bigg]\frac{1}{z^M}.
\label{generatingF}
\end{equation}
We have explicitly written the generating function as a power series for $1/z$ by grouping all terms of order $M$ together by the imposed constraint in the sum over the indices. Let us denote one of these constrained sums 
\begin{equation}
S^{(M)}_{n_t}=\sum_{i_1+i_2+...+i_{n_t}=M}a_1^{i_1}a_2^{i_2}\cdots a_{n_t}^{i_{n_t}},
\end{equation}
by the indices specifying the number of terms $n_t$ and the value of the sum of the indices $M$. We note that this can be obtained from the following contour integral in the complex plane 
\begin{equation}
S^{(M)}_{n_t}=\frac{1}{2\pi \ii}\oint_{\mathcal{C}} z^{M-1} f_{n_t}(z) \mathrm{d}z=\frac{1}{2\pi \ii}\oint_{\mathcal{C}} z^{M+{n_t}-1} \bigg(\frac{1}{z-a_1}\bigg)\cdots\bigg(\frac{1}{z-a_{n_t}}\bigg) \mathrm{d}z
\end{equation}
where $\mathcal{C}$ is a closed contour required to enclose all poles $a_1,\ldots a_{n_t}$. The sum $S^{(M)}_{n_t}$ is the prefactor of $z^{-1}$ in the Laurent series of $z^{M-1}f_{n_t}(z)$.
In the particular case where there are no degenerate values, i.e. $a_k\neq a_{k'}$, for any $k\neq k'$, the sum is evaluated by summing up the residues of simple poles
\begin{equation}
S^{(M)}_{n_t} = \sum_{\text{poles}} \text{Res}\Big[z^{M-1}f_{n_t}(z)\Big]_{z=a_k}=\sum_{k=1}^{{n_t}}\Big[\frac{1}{\prod_{j\neq k}{(a_k-a_j)}}\Big]a_k^{M+{n_t}}.
\label{magicsum}
\end{equation}
Above we used the simple pole residue formula: 
\begin{equation}
\text{Res}\Big[z^{M-1}f_{n_t}(z)\Big]_{z=a_k} = \lim_{z \to a_k}\left[z^{M+{n_t}-1} \left(\frac{1}{z-a_1}\right)\cdots\left(\frac{1}{z-a_k}\right)\cdots\left(\frac{1}{z-a_{n_t}}\right)\left(z - a_k\right)\right].
\end{equation}
Assuming that only one of the values is degenerate $a_k=a_{\bar{k}}$, then the number of terms in the sum is reduced to $n_t-1$, as one of the terms in the generating function appears as $1/(z-a_k)^2$. The residue sum then only goes over $n_t-1$ poles, $n_t-2$ of them being simple poles and the extra one being a double pole, for which the residue formula is:
%%%%%%%%%%%%%%%%
\begin{equation}
    \text{Res} \Big[z^{M-1}f_{n_t}(z)\Big]_{z=a_k=a_{\bar{k}}} = \lim_{z \to a_k} \dv{z} \left[z^{M+{n_t}-1} \left(\frac{1}{z-a_1}\right) \cdots \left(\frac{1}{z-a_k}\right)^2 \cdots \left(\frac{1}{z-a_{n_t}}\right) \left(z-a_k\right)^2 \right].
\end{equation}
%%%%%%%%%%%%%%%%

\section{Details on the Jordan decomposition}
\label{App.:B}
Let us discuss the explicit form of the Jordan decomposition of $H$, Eq.~\eqref{eq:jordandecomp}. As $H$ has at most two-fold degenerate eigenvalues $-h_j$, the associated block in the Jordan normal form $J$ can be either $J_1(-h_j) \oplus J_1(-h_j)$, or $J_2(-h_j)$. In the following, we show the latter is the case by constructing the eigenvectors $\mathbf{v}^{(j)}$ and the generalized eigenvectors $\mathbf{w}^{(j)}$ satisfying
\begin{equation}
    H \cdot \mathbf{v}^{(j)} = -h_j \mathbf{v}^{(j)}, \qquad
    [H - (-h_j) \, \mathbb{1}] \cdot \mathbf{w}^{(j)} = \mathbf{v}^{(j)}. \label{eq:eval}
\end{equation}
Assuming that we have shown the existence of the generalized eigenvectors for $j=n+1,\ldots,N$ with $n = \lceil N/2 \rceil$ (ceiling of $N/2$), the Jordan normal form $J$ of $H$ for even $N$ is given by Eq.~\eqref{eq:Jeven}, and for odd $N$ by
\begin{equation}
    J = J_1(h_n) \oplus J_2(-h_{n+1}) \oplus \cdots \oplus J_2(-h_N) \oplus J_1(0).
\end{equation}
Having determined the eigenvectors and generalized eigenvectors, the similarity matrix $T$ is constructed as
\begin{equation}
    T = \left( \begin{array}{c|c|c|c|c|c|c|c} 
    \mathbf{v}^{(n)} & \mathbf{v}^{(n+1)} & \mathbf{w}^{(n+1)} & \cdots & \mathbf{w}^{(N-1)} & \mathbf{v}^{N} & \mathbf{w}^{N} & \mathbf{v}^{(N+1)} \end{array} \right),
\end{equation}
which contains eigenvectors and generalized eigenvectors as columns, $\mathbf{v}^{(n)}$ exists only for odd $N$, and $\mathbf{v}^{(N+1)}$ is the eigenvector associated with the eigenvalue $-h_0 = -h_{N+1} = 0$, which is convenient for labeling. The technicalities are now the determination of the eigenvectors, generalized eigenvectors, and the determination of $T^{-1}$. \par
Solving Eq.~\eqref{eq:eval} for all $N+1$ components labeled by $m$ yields
\begin{align}
    \mathbf{v}_m^{(j)} &= \prod_{i=j+1}^{\bar{m}} \frac{h_{i-1}}{h_i-h_{j}}, \qquad \text{if } \bar{j} > m\geq 0,\\
    \mathbf{w}_m^{(j)}&=\mathbf{w}_{m}^{\mathrm{u}\,(j)} + \mathbf{w}_{m}^{\mathrm{d}\,(j)},
\end{align}
with
\begin{equation}
    \mathbf{w}_{m}^{\mathrm{u}\,(j)} = \frac{1}{h_{j-1}} \prod_{i=\bar{m}+1}^{j-1} \frac{h_i-h_j}{h_{i-1}}, \quad \text{if } j \geq m > \bar{j}, \quad \text{and} \quad
    \mathbf{w}_{m}^{\mathrm{d}\,(j)} = \prod_{i=j+1}^{\bar{m}} \frac{h_{i-1}}{h_i-h_j}\sum_{i=\bar{m}+1}^{N+1} \frac{1}{h_i-h_j}, \quad  \text{if }\bar{j} \geq m \geq 0,
\end{equation}
and all other components are zero. \par
To determine $T^{-1}$, we permute the columns of $T$ with a permutation matrix $P$ via $T=\tilde{T} \cdot P$, so that
\begin{equation}
    \tilde{T} = \left( \begin{array}{c|c|c|c|c|c|c|c|c} 
    \mathbf{w}^{(N)} & \mathbf{w}^{(N-1)} & \cdots & \mathbf{w}^{(n+1)} & \mathbf{v}^{(n)} & \mathbf{v}^{(n+1)} & \cdots & \mathbf{v}^{(N)} & \mathbf{v}^{(N+1)}
    \end{array} \right).
\end{equation}
In this form, $\tilde{T}$ is a lower triangular matrix. Then, we can write it, and its inverse $\tilde{T}^{-1}$, as
\begin{equation}
    \tilde{T} = \begin{pmatrix}
        \mathbf{T}_{11} & \mathbf{0} \\
        \mathbf{T}_{21} & \mathbf{T}_{22}
    \end{pmatrix},
    \qquad
    \tilde{T}^{-1} = \begin{pmatrix}
        \mathbf{T}_{11}^{-1} & \mathbf{0} \\
        -\mathbf{T}_{22}^{-1}\mathbf{T}_{21}\mathbf{T}_{11}^{-1} & \mathbf{T}_{22}^{-1}
    \end{pmatrix},
\end{equation}
where $\mathbf{T}_{11}$ and $\mathbf{T}_{22}$ are invertible, lower triangular matrices. Then, one can convince oneself that the matrix elements of $\mathbf{T}_{11}^{-1}$ and $\mathbf{T}_{22}^{-1}$ are given by
\begin{align}
    \left(\mathbf{T}_{11}^{-1}\right)_{mj} &= h_m \prod_{i=m+1}^{j} \left(\frac{h_i}{h_i-h_m}\right) \prod_{i=n+1}^{m-1} \left(\frac{h_i}{h_i-h_m}\right)^2 \times
    \begin{cases}
        1, & \text{if $N$ even}, \\
        \dfrac{h_n}{h_n-h_m}, &\text{if $N$ odd},
    \end{cases}\\
    \left(\mathbf{T}_{22}^{-1}\right)_{mj} &= \prod_{i=j}^{\bar{m}-1} \frac{h_i}{h_i-h_{\bar{m}}}.
\end{align}
Finally, the Jordan decomposition of $H$ and the time evolution operator are, respectively, given by
\begin{equation}
    H = \tilde{T} \cdot P \cdot J \cdot P^{-1} \cdot \tilde{T}^{-1}, \qquad \text{and} \qquad U(t) = \exp \left( H \Gamma t \right) = \tilde{T} \cdot P \cdot \exp \left( J \Gamma t \right) \cdot P^{-1} \cdot \tilde{T}^{-1}.
\end{equation}
\section{Details on the Resolvent}
\label{App.:C}
The resolvent $R(z)$ has the matrix elements
\begin{equation}
   R_{mm'}(z) = \frac{1}{z+h_m}\prod_{j=m+1}^{m'} \frac{h_j}{z+h_j},
\end{equation}
which can be shown by directly evaluating \mbox{$R(z) \cdot ( z \, \mathbb{1} - H )=\mathbb{1}$}. With this, the solution for an arbitrary initial condition in Laplace space reads
\begin{equation}
   \bar\rho_m(z) = \sum_{m'=0}^N R_{mm'}(z) \rho_{m'}^{(0)}.
\end{equation}
To transform back to the time domain, the inverse Laplace transform yields
\begin{equation}
    \rho_m(t) 
    = \frac{1}{2 \pi \mathrm{i}}\oint_{\mathcal C} \mathrm{d}z \Big(\sum_{m'} R_{mm'}(z) \, \rho_{m'}^{(0)} \Big) e^{z\Gamma t}
    = \sum_{m'} \left( \frac{1}{2 \pi \mathrm{i}} \oint_{\mathcal{C}_{m'}} \! \! \! \! \! \mathrm{d}z \, R_{mm'}(z)\, e^{z \Gamma t} \right)\rho_{m'}^{(0)},
\end{equation}
where the contour $\mathcal{C}$ encloses all poles of $\sum_{m'} R_{mm'}(z) \rho_{m'}^{(0)}$, and after the second equality, we exchanged the integration and summation, so the contours $\mathcal{C}_{m'}$ enclose all poles of $R_{mm'}(z)$. With $\rho_{m'}^{(0)} = \delta_{m'm_0}$, this reproduces Eq.~\eqref{any-state} and Eq.~\eqref{eq:fm0m} with $z \to -z$ and $h_p \to -h_p$.
\end{document}